# Fuzzy-AHP approach using Normalized Decision Matrix on Tourism Trend Ranking based-on Social Media

Shoffan Saifullah*

Department of Informatics Engineering, Universitas Pembangunan Nasional Veteran Yogyakarta
*shoffans@upnyk.ac.id*
*\* corresponding author*

ABSTRACT

This research discusses multi-criteria decision making (MCDM) using Fuzzy-AHP methods of tourism. The fuzzy-AHP process will rank tourism trends based on data from social media. Social media is one of the channels with the largest source of data input in determining tourism development. The development uses social media interactions based on the facilities visited, including reviews, stories, likes, forums, blogs, and feedback. This experiment aims to prioritize facilities that are the trend of tourism. The priority ranking uses weight criteria and the ranking process. The highest rank is in the attractions of the Park/Picnic Area, with the final weight calculation value of 0.6361. Fuzzy-AHP can rank optimally with an MSE value of ≈0.0002.

## I. Introduction

Tourism is one of the largest sectors in the industrial area in various countries in the world [1]. Tourism is said to be quality if it has a destination features and quality services, as well as the experience felt by someone from the tourist destination visited [2]. Thus that in the field of tourism can be used as a business in the field of services and products [3], [4]. These reasons resulted in an area that would increase the potential for tourism development [5] as well as tourist attractions in East Asia [6]. Improvement can be made by knowing the potential that exists in the region. Factors affecting tourist interest are characteristics of tourism [7].

Based on the existing tourism potential, an area will approach tourists. This approach can be known based on the activities, interests, and opinions of tourists [8]. Thus, tourists will visit tours based on the available facilities and any information about the journey [6]. This information is useful and can be used as a reference for visiting tours.

Along with developing information and communication technology, information about everything can increase via the internet and social media [9]. Most travelers have social media accounts and can access the internet. These connections make them share information about what they are doing (certain content, for example, travel) to social media [10]. Any information published on social media will be considered by other people to travel.

Useful information based on content shared on social media can also be used to improve the quality of tourism potential by an area. Quality improvement will be based on the ranking/rating of existing tours, which will be carried out according to the level of development priority scale. In this case, it refers to data from the internet and social media.

The ranking process will require quite a long time if manually analyzed. So to process data quickly requires a ranking method that is Analytical Hierarchy Process (AHP). AHP will be used in the process of ranking tourism in East Asia to determine the popularity of existing tourism.

Tourism will be the main attraction both in travel and research. Many researchers have conducted various studies on tourism, especially those related to computational methods [6], [11], [12], [13]. Computational approaches (especially in the IT field) include evaluating tourism websites [14], [15], [16].

In addition to the website, tourism development can be done by observing tourists for their response to the tours they visit. Tourists who visit tourist attractions will feel and enjoy it and will share their experiences with social media. Thus, social media can be used to review what tourism is becoming the current trend. Especially in East Asia, many tours are often visited by both domestic and foreign tourists.

The development effort has been researched and carried out by applying several methods to prioritize development based on trends that are of interest to tourists. One way is to use data from social media. Previous studies have carried out specific priority processes using the MCDM method including,

The use of MCDM methods (Fuzzy-AHP and Topsis) in selecting the best places to travel is a matter of decision making based on several criteria that reflect the preferences of travelers [17]. The process of evaluating Smart tourist attraction (STA) is based on the evaluation of tourists using the fuzzy-AHP method and Importance-performance analysis (IPA). This process is used to evaluate and develop the STA Hongshan Zoo, a tourist attraction in China [18].

Application of the integration of case-based reasoning (CBR) and AHP methods build intelligent travel destination planning tools. This method is used to make intelligent decisions by taking the response that best suits the customer. However, rule-based planning cannot be implemented due to unclear customer filtering preferences that can change according to changing customer contexts. Thus, in this study, it is recommended to use the fuzzy method to correct uncertain problems [19].

The structure of this article is organized into several sections, namely: the second part explains the tourism studies that have been done before. Section 3 describes the research methodology, such as data, tools, and methods (AHP) used. Section 4 explains experiments and testing in detail (results and discussion). The last part (in section 5) explains the conclusions of the research that has been done.

## II. Method

In this study, we secured a data set from the UCI Machine Learning Library website. This research will rank tourism data based on reviews of 10 destinations in 10 categories in East Asia. This data was obtained from user ratings taken from social media (data set is populated by crawling TripAdvisor.com [6]). The data used were 980 social media user data that provide an assessment of the tourist attractions visited (based on data from social media). Social media users offer a range of 0 to 4 (as in Fig. 1).

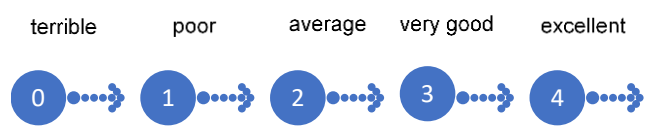

Fig. 1. Range of tourist rating categories for tourist destinations

Each traveler gives an assessment that will be categorized in several criteria, namely Excellent (4), Very Good (3), Average (2), Poor (1), and Terrible (0), as shown in Fig. 1. The assessment results will be averaged, and the average rating is used for each category per user.

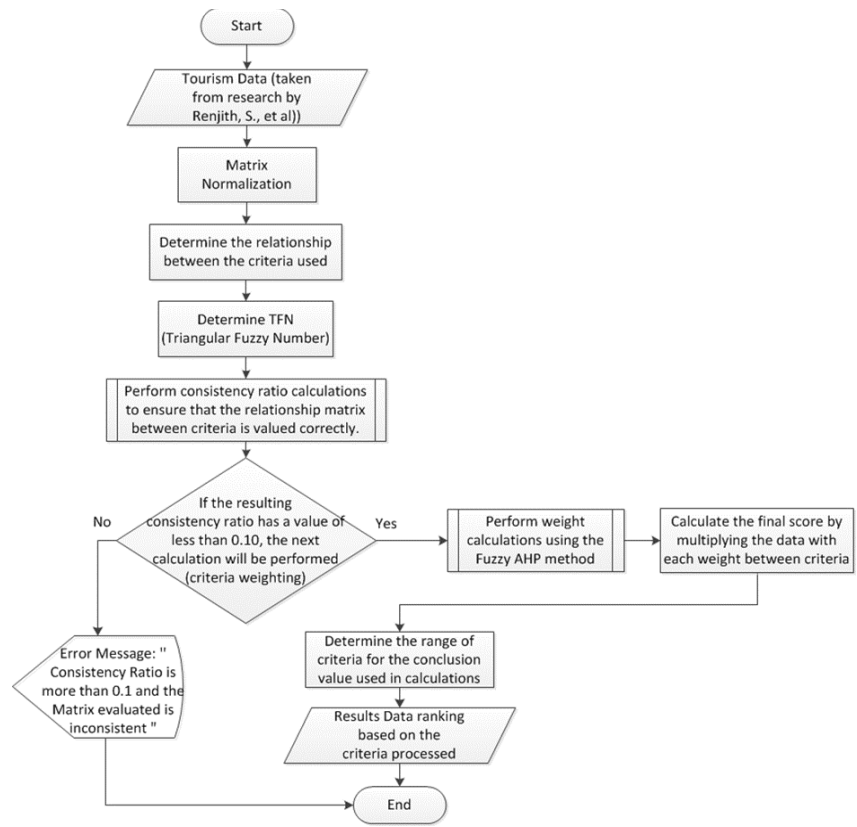

Fig. 2. Flowchart Fuzzy AHP method in ranking tourist attractions based on social media data

Based on the data acquisition, the average results will be ranked using the fuzzy-AHP method. The flowchart in Fig. 2 shows the ranking process with the fuzzy-AHP. Based on the flowchart (Fig. 2), the initial step is to enter the average data from tourist ratings based on the use of social media into the system. The data is read as a matrix and normalized. The process is used to get a small value so that the calculation is maximum. This matrix normalization is used to align various values of the attribute. The normalization process uses the equation formula (1).

$$r_{ii} = \frac{x_{ij}}{max_i\, x_{ij}}; j = 1, 2, 3, \dots, m; i = 1, 2, 3, \dots, n \qquad (1)$$

In this equation, $r_{ij}$ is a normalized matrix, $x_{ij}$ is a decision matrix, and $max_i$ is the maximum value of each column. This normalization process uses the simple addictive weight (SAW) normalization equation [20].

The normalization matrix will be used in the process of determining relations between criteria. Relationships between criteria will be made using the analytic hierarchy process (AHP) method. AHP is used to find the ratio scale based on the comparison of criteria pairs [21]. AHP can handle the problem of quantitative and qualitative decision making for simple problems [22]. The comparison criteria scale used in the AHP Method uses the evaluation scale suggested by Saaty, divided into ranges of 1 to 9 points.

In several studies, AHP was developed with the addition of methods, one of which is the combination of Fuzzy and AHP. Fuzzy-AHP is a method of developing AHP that can handle fuzzy decisions [23], [24]. Fuzzy used is TFN (Triangular Fuzzy Number), a set of 3 numbers that form fuzzy graphs at fuzzy values of 0, then go up to 1, and return to 0 [25]. TFN contains 2 groups of data, of which the first group is TFN in the actual value, and the second group is the inverse of TFN, namely by changing x to 1/x and reversing the order of the TFN numbers.

In the AHP method, the decision-making process makes it easy for humans to make decisions. These decisions are that humans when making decisions using the concept of perception to allow inconsistencies to occur. AHP will decide whether the perception is consistent or not. This concept of consistency is formulated in a consistency index (2).

$$CI = \frac{t-n}{n-1} \quad (2)$$

CI is the consistency index, t is the most considerable normalization value of an ordered matrix n, and n is the order matrix. This CI calculation is used to check the consistency of the pairwise comparison matrix of Saaty. This matrix is said to be consistent if the value of CI is zero (0). Saaty also stated that inconsistency has limits set by using a Consistency Ratio (3).

$$CR = \frac{CI}{IR} \quad (3)$$

Consistency Ratio (CR) is a comparison between the consistency index (CI) with a random index value (IR). The maximum CR limit is 0.1 (CR ≤ 0.1). This condition states that inconsistencies in decision making are still accepted, and if not suitable, they will be reprocessed.

Acceptance of the decision-making process based on CR values will determine the weight calculation process. The method used in this calculation is Fuzzy AHP. The process includes two main steps, namely converting each relation between criteria into a Triangular Fuzzy Number (TFN) and calculating the degree of probability.

The initial process of Fuzzy AHP, calculates the relation matrix between criteria, by entering the value in the lower triangle according to the value in the relationship between criteria. If the relation matrix value between criteria is more than 1, then the TFN value used is the criterion value in the first group, in addition to using the criterion value in the second group. After that, the values of each TFN used in the matrix between criteria are added up, the results of which are used to add values to each criterion. Besides, the calculation of each value is divided by the number of relations in each column.

The estimation of the weight value can be obtained by calculating the degree of probability. This process considers the principle of comparison between fuzzy numbers. This comparison is obtained by determining the value of the vector (V) [26].

$$V(M2 \geq M1) = \begin{cases} 1 & , if\ m_2 \geq m_1 \\ 0 & , If\ l_1 \geq u_2 \\ \frac{(l_1 - u_2)}{(m_2 - u_2) - (m_1 - l_1)} & , other \end{cases} \quad (4)$$

The possible degrees are obtained based on the calculation of each data with row and column indexes. The results are converted into vector equation (4). The minimum possible degree value of each criterion will be weighted and normalized by dividing each weight value by all the resulting weights. The end of the process is determining the range of criteria for the final calculation and ranking for all data.

The final result of normalized weighting is calculated the Mean Squared Error value to determine the accuracy of the ranking method. Calculation of Mean Squared Error (MSE) represents the quadratic error value (5).

$$MSE = \frac{1}{n}\sum_{i=1}^{n}(f_i - y_i)^2 \quad (5)$$

### III. Result and Discussion

In this section, we will discuss the results of the research including the following.

*A. Data Analysis*

The data used is data in the form of a matrix measuring 10 x 980 data. The data is carried out the normalization process using the normalization equation formula (1). So that when it is made visually, as in Fig. 3.

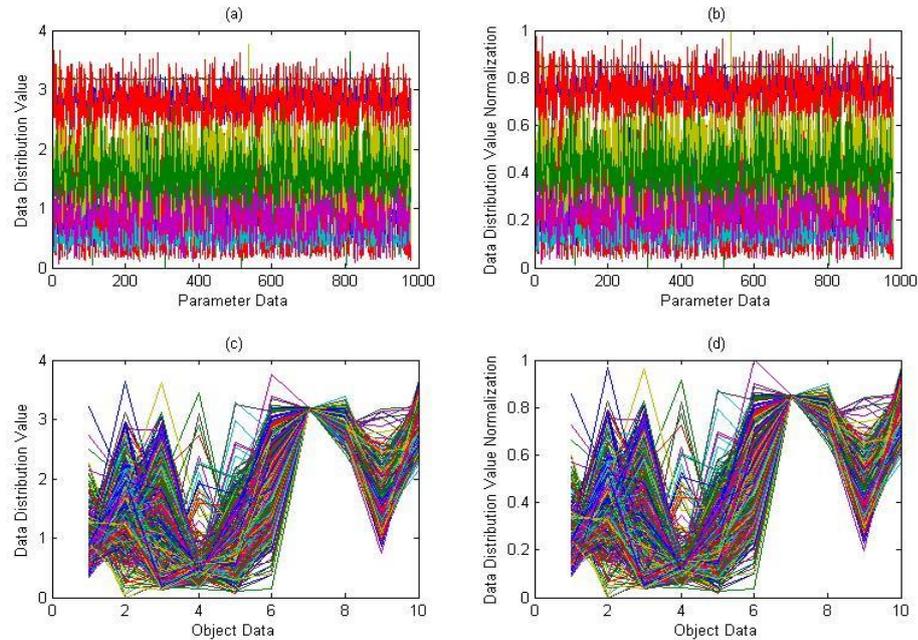

Fig. 3. Visualization of the distribution of social media data on the assessment of tourism groups in East Asia based on parameters and real objects (a, c) and normalized (b, d)

Fig. 3 shows the distribution of data and parameters of tourism destinations in East Asia. The complete and real initial data and parameters are shown in Fig. 3(a) for parameters and Fig. 3(c) for object data used for weight calculation. When this initial data is processed directly, it requires calculations that produce less than the maximum value, so the normalization process needs to be done. The results of the normalization process are shown in Fig. 3(c) and Fig. 3(d). Normalized graphs shown have a maximum data value of 1.

### B. Relationship between Criteria and Use of the Triangular Fuzzy Number

Relationships between criteria are made based on parameter data. The results of relations between these criteria will be used using TFN. Thus, the resulting TFN is shown in Table 1.

Table 1. Comparative Scale of Criteria

| Intensity of Interest AHP | Linguistic labels | TFN | |
|---|---|---|---|
| | | *Real* | *Inverse* |
| 1 | Just Equal | 1 1 1 | 1 1 1 |
| 2 | Intermediate | $\frac{1}{2}\ \frac{3}{4}\ 1$ | $1\ \frac{4}{3}\ 2$ |
| 3 | Moderately Important | $\frac{2}{3}\ 1\ \frac{3}{2}$ | $\frac{2}{3}\ 1\ \frac{3}{2}$ |
| 4 | Intermediate | $1\ \frac{3}{2}\ 2$ | $\frac{1}{2}\ \frac{2}{3}\ 1$ |
| 5 | Strong Important | $\frac{3}{2}\ 2\ \frac{5}{2}$ | $\frac{2}{5}\ \frac{1}{2}\ \frac{2}{3}$ |
| 6 | Intermediate | $2\ \frac{5}{2}\ 3$ | $\frac{1}{3}\ \frac{2}{5}\ \frac{1}{2}$ |
| 7 | Very Strong | $\frac{5}{2}\ 3\ \frac{7}{2}$ | $\frac{2}{7}\ \frac{1}{3}\ \frac{2}{5}$ |
| 8 | Intermediate | $3\ \frac{7}{2}\ 4$ | $\frac{1}{4}\ \frac{2}{7}\ \frac{1}{3}$ |
| 9 | Extremely Strong | $\frac{7}{2}\ 4\ \frac{9}{2}$ | $\frac{2}{9}\ \frac{1}{4}\ \frac{2}{7}$ |

Table 1 shows the intersection of AHP interests with the relationship between the criteria used in the TFN. Thus, the TFN value used is the two TFN groups. TFN is used in 2 groups of data, of which the first group is TFN in the real value, and the second group is the inverse (1 value / true value and reverses the order of the TFN number).

TFN will be used in calculations if the consistency ratio value resulting from the consistency calculation of AHP is less than 0.1. AHP consistency calculation is done based on randomly predetermined indexes, several criteria, lambda values, and consistency index. In the experiments conducted, the results obtained in Table 2.

Table 2. Comparative Scale of Criteria

| Result Variable | Result Value |
|---|---|
| **Random Index** | range 0 s/d 180 |
| **Lambda Max.** | 1 |
| **Average weight between criteria** | 0.001 |
| **Consistency Index** | -1 |
| **Consistency Ratio** | -0.0056 |

The calculation of the Consistency Ratio, besides being influenced by the consistency index, is also strongly affected by the random range of the index. The higher the random index range used in the consistency ratio, the results of the consistency ratio close to 0 (zero).

Based on Table 2, the consistency ratio value is -0.0056, indicating that the condition of the consistency ratio is less than 0.1. Thus, the next process is quantified fuzzy AHP.

Fuzzy AHP calculation uses the TFN table results (Table 1) and some of the required components, including the degree of likelihood produced, the relation per row of the criteria, the maximum weight used. Parts other than TFN needed for the calculation of Fuzzy AHP are shown in Table 3.

Table 3. Variables In Calculation of Fuzzy AHP

| Variables | Values |
|---|---|
| **Average degree of probability** | 327.33 |
| **Average number of relations** | 0.001 |
| **Weight Max.** | 1 |

The results of the AHP consistency calculation are used to produce output weight values. This weight will be used to multiply the initial data. Thus, the results of the multiplication between the data and the weight output will be ranked.

*C. Ranking of Fuzzy AHP methods*

Based on the multiplication of the calculation weight of Fuzzy AHP and its data, the results obtained as in Fig. 4.

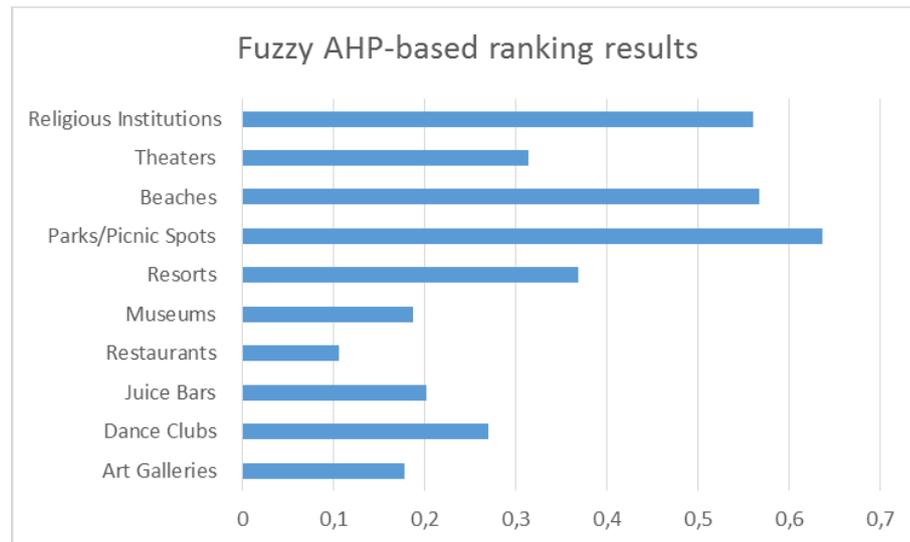

Fig. 4. Fuzzy-AHP calculation results at each tourist site based on users' feedback

Based on Fig. 4, the ranking process is done by decreasing the value in Ascending. Ranking results show that parks/picnic spots give the most considerable amount of 0.6361. Each tourist destination has been calculated and ranked so that the results obtained as in Table 4.

Table 4. Table 4. The Ranking of Tourist Attractions Using The Fuzzy-AHP Method on Social Media Data

| Rank | Average user feedback on and the results tourist attractions | Fuzzy AHP Calculation | | Manual calculation | |
|---|---|---|---|---|---|
| | | 1[a] | 2[b] | 1[a] | 2[b] |
| 1 | Parks/Picnic Spots | 0.6361 | 0.0555 | 3.1809 | 0.8459 |
| 2 | Beaches | 0.5670 | 0.0743 | 2.8350 | 0.7540 |
| 3 | Religious Institutions | 0.5598 | 0.0560 | 2.7992 | 0.7444 |
| 4 | Resorts | 0.3685 | 0.0310 | 1.8428 | 0.4901 |
| 5 | Theaters | 0.3138 | 0.0570 | 1.5694 | 0.4174 |
| 6 | Dance Clubs | 0.2705 | 0.0980 | 1.3526 | 0.3597 |
| 7 | Juice Bars | 0.2026 | 0.1982 | 1.0133 | 0.2694 |
| 8 | Museums | 0.1879 | 0.1673 | 0.9397 | 0.2499 |
| 9 | Art Galleries | 0.1786 | 0.0990 | 0.8931 | 0.2375 |
| 10 | Restaurants | 0.1065 | 0.1530 | 0.5325 | 0.1416 |

[a] Data calculation with real value
[b] Normalized data calculation first

Based on Table 4, the calculation results from the AHP fuzzy method, and the manual can be checked for accuracy. The level of accuracy in this study was calculated using the MSE formula (1). MSE calculations give an average value of all data of $\approx 0.0002$. The results of this MSE show that the fuzzy-AHP method can be used to rank tourist destinations in East Asia. It is because the MSE value generated is minimal.

## IV. Conclusion

This study combines fuzzy and AHP methods to rank tourism trends in East Asia. Based on the experiments conducted, the AHP fuzzy method can contribute to ranking with a success rate of 100% with an MSE value of $\approx 0.0002$ for calculation data based on real calculations and normalization results.

## Acknowledgment

The author is grateful for the supporting research of Universitas Pembangunan Nasional "Veteran" Yogyakarta. And especially thankful for the Department of Informatics Engineering had allowed and supported to publish this article.